\documentclass{appolb}
\usepackage{graphicx}

\begin{document}
\title{QCD Phase Diagram at NICA energies:\\
$K^+/\pi^+$ horn effect and light clusters in THESEUS%
\thanks{Presented at NICA Days and 4th MPD Collaboration Meeting, Warsaw, 21.10.2019}%
}
\author{D. Blaschke$^{a,b,c}$, A.V. Friesen$^{a}$, Yu.B. Ivanov$^{a,b}$, \\
Yu.L. Kalinovsky$^{a}$, M. Kozhevnikova$^{a}$, S. Liebing$^{a}$, \\
A. Radzhabov$^{d,e}$, G. R\"opke$^{b,f}$
\address{$^{a}$JINR Dubna, 141980 Dubna, Russia
\\
$^{b}$National Research Nuclear University (MEPhI), 115409 Moscow, Russia
\\
$^{c}$University of Wroclaw, 50-204 Wroclaw, Poland
\\
$^{d}$Matrosov Institute System Dynamics \& Control Theory, 664003 Irkutsk, Russia
\\
$^{e}$Irkutsk State University, 664003 Irkutsk, Russia
\\
$^{f}$University of Rostock, 
D-18051 Rostock, Germany}
}
\maketitle
\begin{abstract}
We discuss recent progress in the development of the three-fluid hydro-\-dynamics-based program THESEUS 
towards an event generator suitable for applications to heavy-ion collisions at the intermediate 
energies of the planned NICA and FAIR experiments.
We follow the strategy that modifications of particle distributions at the freeze-out surface in the QCD phase diagram may be mapped directly to the observable ones within a sudden freeze-out scheme. 
We report first results of these investigations for the production of light clusters (deuterons and tritons) which can be compared to experimental data from the HADES and the NA49 experiment and for the interpretation of the "horn" effect observed in the 
collision energy dependence of the $K^+/\pi^+$ ratio.
Medium effects on light cluster production in the QCD phase diagram are negligible at the highest NICA energies but shall play a dominant role at the lowest energies. 
A sharp "horn"-type signal in the  $K^+/\pi^+$ ratio can be obtained when the onset of Bose condensation modelled by a pion chemical potential results in an enhancement of pions at low momenta (which is seen at LHC energies) and would occur already in the NICA energy range. 
\end{abstract}
  
\section{Chemical freeze-out and light clusters in the phase diagram}
\begin{figure}[!h]
\includegraphics[width=12.5cm]{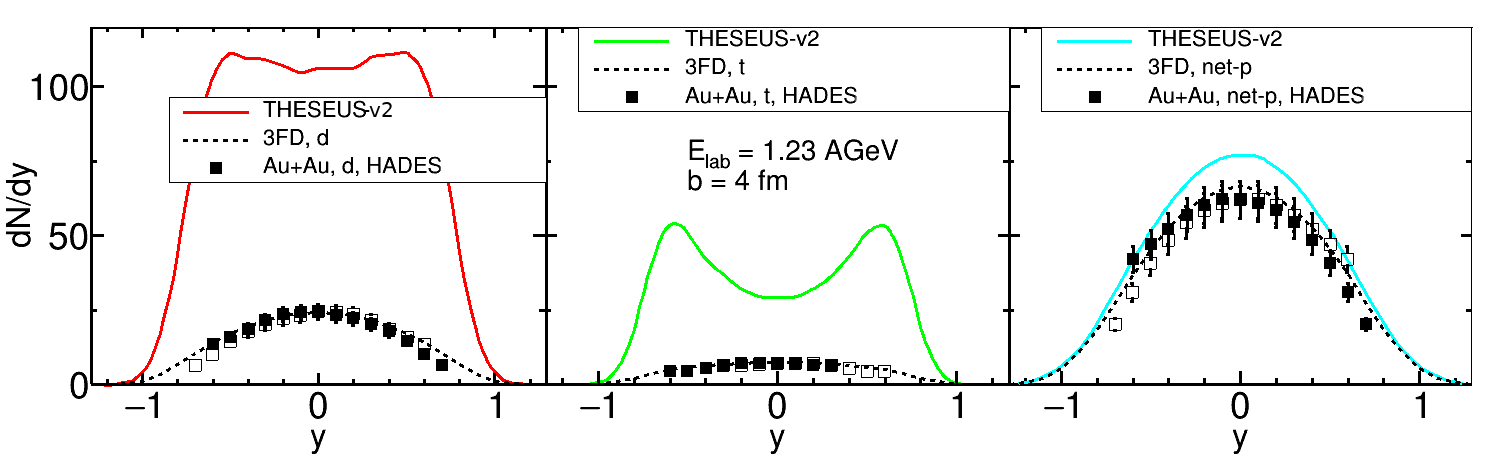}
\caption{Preliminary results of the HADES collaboration \cite{Szala:2019} for the rapidity distribution of deuterons (left panel), tritons (middle panel) and protons (right panel)  in Au+Au collisions at $E_{\rm lab}=1.23$ A GeV (open symbols: reflected data) compared with results from the three-fluid hydrodynamics simulation (THESEUS) \cite{Batyuk:2016qmb,Blaschke:2020gqr} for impact parameter $b=4$ fm using a crossover equation of state model with cluster abundances described by a thermal statistical model without selfenergy effects (solid lines) and the 3FD model \cite{Ivanov:2017nae} with coalescence from \cite{Russkikh:1993ct} (dashed lines). 
}
\label{Fig:HADES}
\end{figure}

In this contribution, we present recent progress in the development of the three-fluid hydro-\-dynamics-based program THESEUS
\cite{Batyuk:2016qmb}  towards an event generator suitable for applications to heavy-ion collisions at the intermediate energies of the planned NICA and FAIR experiments. We follow the strategy that modifications of particle distributions at the freeze-out surface in the QCD phase diagram may be mapped directly to the observable ones within a sudden freeze-out scheme. 
%
\vspace{-1cm}
\begin{figure}[!h]
\includegraphics[width=12.5cm]{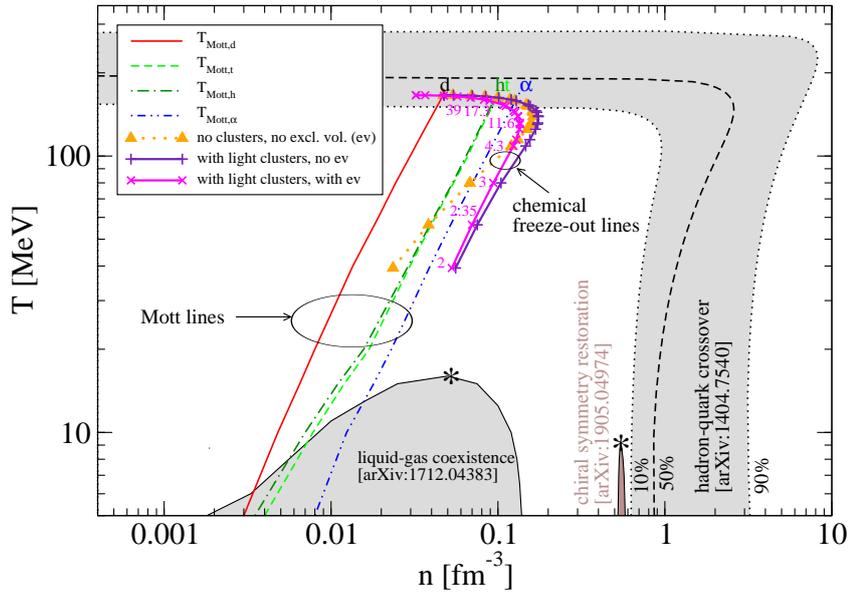}
\caption{Phase diagram in the temperature-density plane including effects of light clusters. 
The yellow line is chemical freeze-out without clusters from \cite{Randrup:2009ch}. 
Chemical freeze-out and Mott lines for light clusters are from \cite{Blaschke:2020gqr}. 
The coexistence region for the nuclear gas-liquid transition is shown as grey shaded region according to \cite{Typel:2017vif}
with its critical point (star) in top.
The hadron-quark matter transition is described as a crossover with a switch function from \cite{Albright:2014gva} defining the lines where the quark matter partial pressure amounts to $10\%$, $50\%$ 
and $90\%$ of the total pressure, respectively.
The brown filled region indicates a first-order chiral symmetry restoration transition by parity doubling in hadronic matter according to \cite{Marczenko:2019trv}.
}
\label{Fig:PhD}
\end{figure}
In Fig.~\ref{Fig:PhD} we show that in comparison to the thermal statistical model without clusters \cite{Randrup:2009ch}
the inclusion of light clusters increases the freeze-out density (evaluated along the freeze-out line in the $T-\mu$ plane 
described by the fit of \cite{Cleymans:2005xv}) by a factor of two \cite{Blaschke:2020gqr} for lowest available collision energies.
We also show the Mott-lines for the light clusters defined by the vanishing binding energy of the clusters at rest in the medium.
At finite cluster momentum relative to the medium these lines would get shifted to higher densities \cite{Schuck:1999eg}.
They are indicators for the region in the phase diagram where in-medium effects on cluster formation at the freeze-out have to be taken into account \cite{Roepke:2017ohb}.
In Fig.~\ref{Fig:HADES} we demonstrate by comparing with preliminary HADES data \cite{Szala:2019} for $E_{\rm lab}=1.23$ A GeV that at lowest NICA energies the description of light clusters requires the account for in-medium effects 
\cite{Roepke:2017ohb,Typel:2009sy} while already at energies above $E_{\rm lab}\approx 20$ A GeV an acceptable description is achieved without them \cite{Blaschke:2020gqr}.

\section{Beth-Uhlenbeck approach to the "horn" effect}
\begin{figure}[!h]
\includegraphics[width=6.5cm]{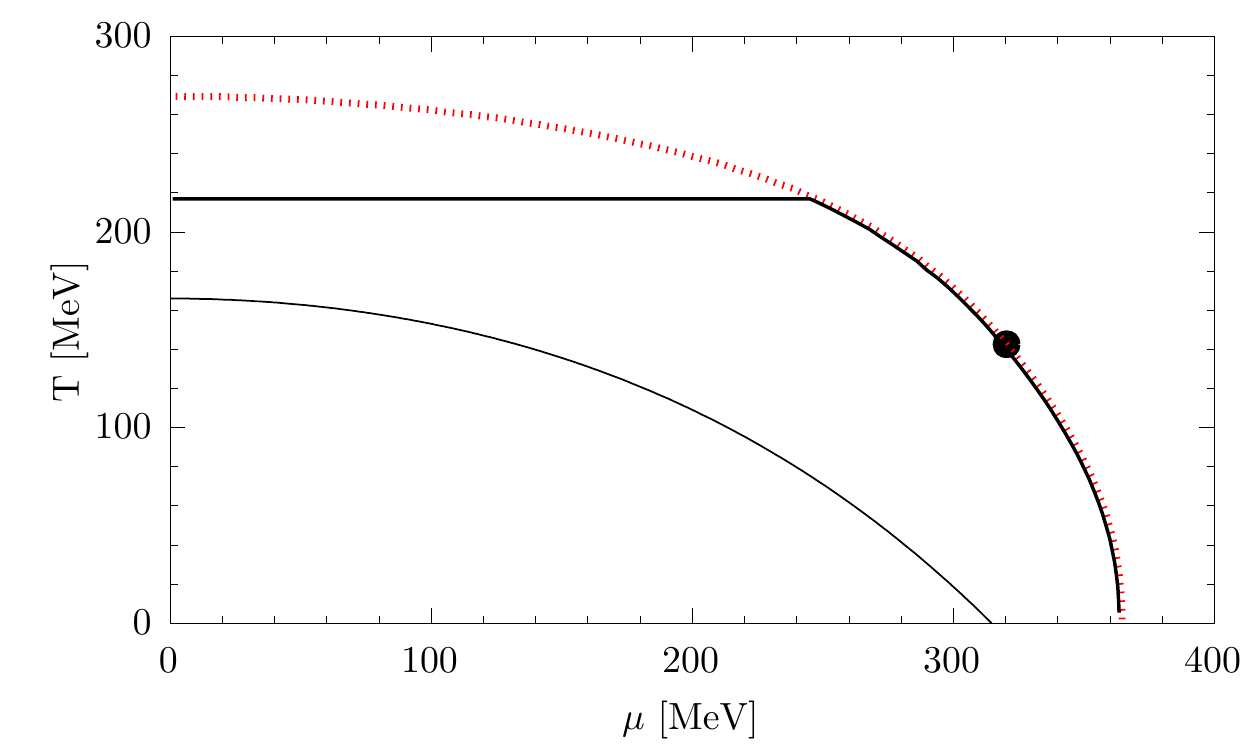}
\includegraphics[width=6.5cm]{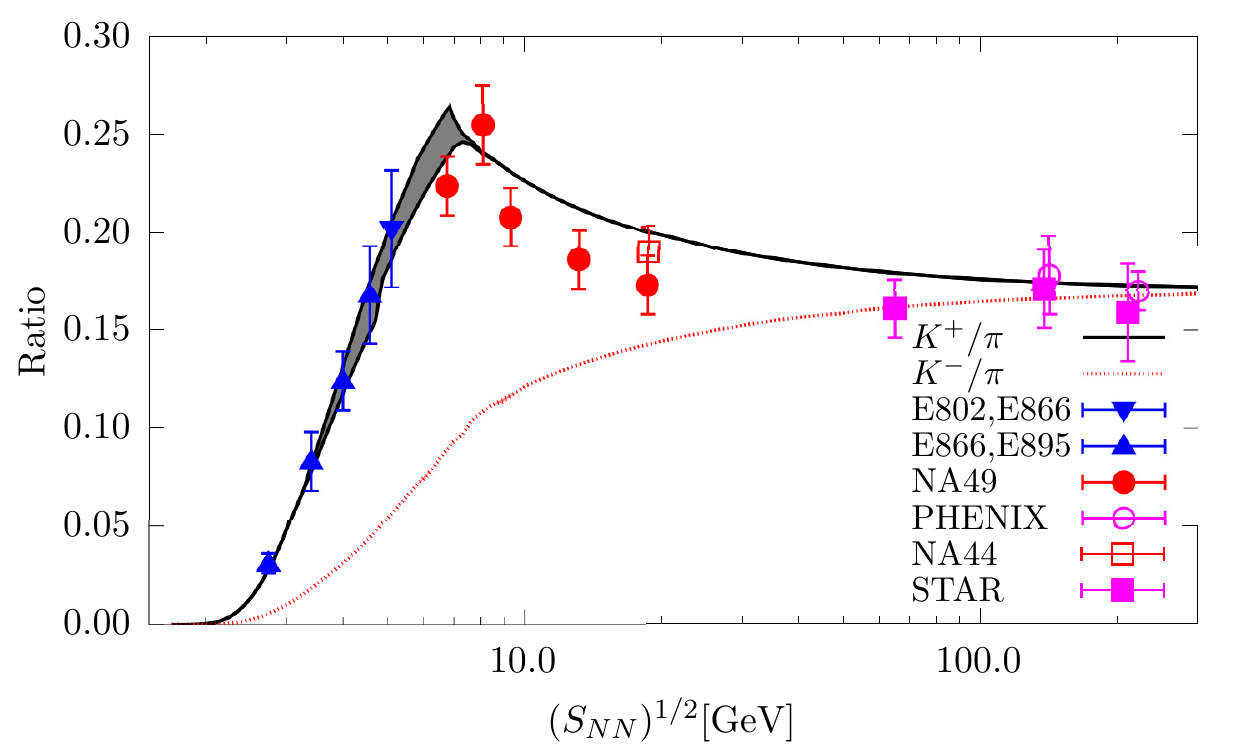}
\caption{Left panel: Phase diagram of the PNJL model and lines for a scan of the $K/\pi$ ratios. 
The red dotted line corresponds to the first order phase transition or crossover transition and the black point denotes
the critical end point. The green dashed line is the Mott temperature for the pion, lines of scan of the $K/\pi$ ratios in the $T-\mu$ plane are: chemical freeze-out \cite{Cleymans:2005xv} (thin black), critical line (red dotted),
critical line+straight part for $\mu_\pi = 134.5$ MeV (thick black).
Right panel: Plot of the $K^+/\pi^+$ and $K^-/\pi^-$ ratios along the scan line defined by the phase transition 
and a straight constant temperature line. 
The shaded region between the lines for $K^+/\pi^+$ corresponds to the contribution of the anomalous mode,
 see Ref.\cite{Blaschke:2019col}.}
\label{Fig:Kpi}
\end{figure}

The Beth-Uhlenbeck expression for the ratio of the yields of kaons and pions is defined as ratio of their partial number densities 
\cite{Dubinin:2016wvt}
\begin{eqnarray}
\label{K+pi+n-ratio}
\frac{n_{K^\pm}}{n_{\pi^\pm}} = 
\frac{\int dM \int d^3p\  (M/E)g_{K^\pm}(E)[1+g_{K^\pm}(E)]\delta_{K^\pm}(M)}
{\int dM \int d^3p\ (M/E)g_{\pi^\pm}(E)[1+g_{\pi^\pm}(E)]\delta_{\pi^\pm}(M)}\ , 
\end{eqnarray}
where $E = \sqrt{{M}^2+p^2}$, $g(E) = (e^{E/T}-1)^{-1}$ is the Bose function and $\delta_{M}(M)$ is in-medium phase shift
in the meson channel $M$ calculated within the PNJL model \cite{Dubinin:2016wvt,Blaschke:2019col}.  
In order to relate the model results with the actual phenomenology of chemical freeze-out in heavy-ion collisions 
we use the idea to map points with a fixed value of $\mu/T$ on the line in phase diagram of our PNJL model to points on the curve fitted to statistical model analyses.
For the description of experimental data the scan line is chosen as a critical line in PNJL model supplemented by a straight line at some fixed $T$. This ansatz is needed due to fact that the pseudocritical temperature at zero chemical potential in the PNJL model is too high when compared with lattice QCD and with the fit of the freeze-out line. 
A sharp "horn"-type signal in the  $K^+/\pi^+$ ratio  \cite{Gazdzicki:1998vd} can be obtained when the onset of Bose condensation modelled by a nonequilibrium pion chemical potential \cite{Blaschke:2020afk}
results in an enhancement of pions at low momenta (which is seen at LHC energies \cite{Begun:2015ifa}) and would occur already in the NICA energy range, see Fig.~\ref{Fig:Kpi} and Ref.~\cite{Blaschke:2019col}.

\section{Conclusion}
We have discuss recent progress in the development of the three-fluid hydro-\-dynamics-based program THESEUS 
towards an event generator suitable for applications to heavy-ion collisions at the intermediate 
energies of the planned NICA and FAIR experiments.
We have followed the strategy that modifications of particle distributions at the freeze-out surface in the QCD phase diagram may be mapped directly to the observable ones within a sudden freeze-out scheme. 
As first results of these investigations we have reported on the production of light clusters (deuterons and tritons) which can be compared to experimental data from the HADES and the NA49 experiment as well as on the "horn" effect observed in the 
collision energy dependence of the $K^+/\pi^+$ ratio.
Medium effects on light cluster production in the QCD phase diagram are negligible at the highest NICA energies but shall play a dominant role at the lowest energies. 
A sharp "horn"-type signal in the  $K^+/\pi^+$ ratio can be obtained when the onset of Bose condensation modelled by a pion chemical potential results in an enhancement of pions at low momenta and would occur already in the NICA energy range. 

As a next step the in-medium effects on cluster and hadron distribution functions shall be included to the particlization routine of THESEUS to improve the description of light cluster formation and the "horn" effects.

\subsection*{Acknowledgements}
The work on the light cluster production was supported by the Russian Science Foundation grant no. 
17-12-01427, and the work on the "horn" effect by the Russian Fund for Basic Research 
grant no. 18-02-40137.

\end{document}